\documentclass[11pt]{article}
\usepackage{setspace}
\usepackage[caption=false]{subfig} %% the caption=false ensures justified figure captions 
\usepackage{graphicx,caption}
\usepackage{authblk} 
\usepackage{indentfirst}
\usepackage[left=3cm,right=2cm,top=2.5cm,bottom=2cm]{geometry}
\usepackage{float}
\usepackage{array}
\usepackage{xcolor}
\usepackage{mathtools}
\usepackage{amsmath,amsfonts}
\usepackage{bbm}
\usepackage{xspace}
\usepackage[font={footnotesize}]{caption}

\usepackage{textcomp}
\usepackage{hyperref}
\usepackage[capitalise,noabbrev]{cleveref}
\crefname{appendixfigure}{Supplementary Figure}{SI Figures}
\usepackage{todonotes}
\usepackage[utf8]{inputenc}
\usepackage{siunitx}
\usepackage{bm}
\usepackage{float}

\usetikzlibrary{patterns}
\newcommand*{\TitleFont}{%
      \usefont{\encodingdefault}{\rmdefault}{b}{n}%
      \fontsize{16}{20}%
     \selectfont}
\usepackage[percent]{overpic}

\makeatletter
\renewcommand{\maketitle}{\bgroup\setlength{\parindent}{0pt}
\begin{flushleft}
 \fontsize{16}{32}
 \textbf{\@title}
 
   \fontsize{12}{24}
  \@author
\end{flushleft}\egroup
}
\newfam\bboardfam
\font\tenbboard=msbm10  
 \font\sevenbboard=msbm7
   \font\fivebboard=msbm5 
\textfont\bboardfam=\tenbboard
\scriptfont\bboardfam=\sevenbboard
\scriptscriptfont\bboardfam=\fivebboard

\newif\ifIncludeFigures
\IncludeFigurestrue % or
\begin{document}
\author[1,*]{Jialong Jiang}
\author[1,2*]{Sisi Chen}
\author[1,2]{Tiffany Tsou}
\author[3]{Christopher S. McGinnis}
\author[1]{Tahmineh Khazaei}
\author[3]{Qin Zhu}
\author[1,2]{Jong H. Park}
%\author[1,2]{Paul Rivaud}
\author[1]{Inna-Marie Strazhnik}
\author[5,6]{Eric D. Chow}
\author[4]{David A. Sivak}
\author[3,7,8,9]{Zev J. Gartner}
\author[1,2\dag]{Matt Thomson}

% Therapeutic algebra of immunomodulation at single-cell resolution
% Therapeutic algebra of human immune cells at single cell resolution
\title{\TitleFont Therapeutic algebra of immunomodulatory drug responses at single-cell resolution}

\affil[1]{Division of Biology and Biological Engineering, California Institute of Technology, Pasadena, California, 91125, USA.}
\affil[2]{Beckman Single-Cell Profiling and Engineering Center, California Institute of Technology, Pasadena, CA, 91125, USA.}
\affil[3]{Department of Pharmaceutical Chemistry, University of California San Francisco, San Francisco, CA, 94143, USA}
\affil[4]{Department of Physics, Simon Fraser University, Burnaby, BC V5A 1S6, Canada}
\affil[5]{Department of Biochemistry and Biophysics, University of California San Francisco, San Francisco, CA, 94143, USA} 
\affil[6]{Center for Advanced Technology, University of California San Francisco, San Francisco, CA, 94143, USA}
\affil[7]{Helen Diller Family Comprehensive Cancer Center, San Francisco, CA, 94115, USA}
\affil[8]{Chan Zuckerberg BioHub, University of California San Francisco, San Francisco, CA, 94143, USA}
\affil[9]{Center for Cellular Construction, University of California San Francisco, San Francisco, CA, 94143, USA}

\affil[*]{These authors contributed equally to this work}
\affil[$\dag$]{mthomson@caltech.edu}
\maketitle

\newpage
\subsection*{Abstract} 
\noindent
\textbf{Therapeutic modulation of immune state is central for the treatment of human disease. However, how drugs and drug combinations impact the diverse cell types in the human immune system remains poorly understood at the transcriptome scale. Here, we apply single-cell mRNA-seq to profile the response of human immune cells to 502 immunomodulatory drugs alone and in combination. We develop a unified mathematical model that quantitatively describes the transcriptome scale response of myeloid and lymphoid cell-types to individual drugs and drug combinations through a single inferred regulatory network. The mathematical model reveals how drug combinations generate novel, macrophage and T-cell states by recruiting combinations of gene expression programs through both additive and non-additive drug interactions. A simplified drug response algebra allows us to predict the continuous modulation of immune cell populations between activated, resting and hyper-inhibited states through combinatorial drug dose titrations. Our results suggest that transcriptome-scale mathematical models could enable the design of therapeutic strategies for programming the human immune system using combinations of therapeutics.}

%\subsection*{Abstract 2} 
%\noindent
%\textbf{Therapeutic modulation of immune state is central for the treatment of human disease. However, how drugs and drug combinations impact the diverse cell types in the human immune system remains poorly understood at the transcriptome scale.  Here, we apply single-cell mRNA-seq to profile the response of human immune cells to 502 immunomodulatory drugs alone and in combination. We develop a unified mathematical model that quantitatively describes the transcriptome scale response of myeloid and lymphoid cell-types to individual drugs and drug combinations through a single inferred regulatory network. While drugs switch T-cells discretely between resting and activated states, the mathematical model reveals how drug combinations generate novel,  macrophage states by recruiting combinations of gene expression programs through both additive and non-additive drug interactions.  A simplified drug response algebra allows us to predict the continuous modulation of the immune cell populations between activated, resting and hyper-inhibited states through combinatorial drug dose titrations. Our results suggest that transcriptome-scale mathematical models could enable the design of therapeutic strategies for programming the human immune system using combinations of therapeutics.}

\section*{Introduction}

The rational programming of the human immune system is a major goal for the treatment of diseases including cancer, auto-immunity, and neurodegeneration~\cite{de2006paradoxical,marrack2001autoimmune,labzin2018innate}. The human immune system is composed of a diversity of cell-types (T-cells, macrophages, and B-cells), and each cell-type adopts specialized sub-states of signaling and gene expression in response to environmental cues. In diseases including cancer and auto-immunity individual cell-types including macrophages and T-cells can become locked into aberrant states of immune suppression or hyper-activation that actually drive disease progression. Signaling interactions between cell types reinforce aberrant immune states and generate a major obstacle for the treatment of conditions like hyper-inflammation and immuno-therapy resistance cancer. An ability to reprogram immune cell populations between states of activation and suppression using combinations of drugs and other therapeutics could allow the rational design of therapeutic strategies for conditions that currently defy treatment. However, immune state programming remains challenging because we lack a detailed understanding of how drugs and other therapeutics act on the large diversity of different cell-states that execute immune function. 

Single-cell genomics methods now enable full transcriptome profiling of heterogeneous cell-populations across large numbers of therapeutic interventions~\cite{Adamson2016-ge, Dixit2016-ol,Subramanian2017-jy,McFarland2020-aa,Srivatsan2020-kn, replogle2022mapping,Burkhardt2021-ob,Chen2018-qc}. Single-cell mRNA-seq measurements can quantify how a drug alters the transcriptional state of individual cell-types, and also reveal how drug action across many cell-types rebalances the abundance of different cell-states in an immune population. However, conventional single-cell data analysis methods typically focus on cell-state identification and differential gene expression and do not allow the construction of quantitative and predictive models that will be required for rational therapeutic design. Immune state programming will require data modeling frameworks that can dissect and predict therapeutic responses in a heterogeneous cell population across thousands of expressed genes cell-types. 

Conceptually, quantitative models might reveal principles that simplify the biological interpretation of combinatorial drug responses while also aiding rational treatment design. For example, previous work has demonstrated that combinations of drugs can induce additive responses on protein expression and dynamics ~\cite{geva2010protein} where additivity means that the response of a specific gene to a drug combination can be predicted as a simple linear combination of single drug effects. The concept of drug additivity has also enabled the identification of non-additive drug responses \cite{nichols2011phenotypic, yeh2006functional}. Non-additive drug responses are associated with interactions between pathways targeted by individual drugs, and therefore, reveal points of information integration in cellular regulatory networks. While principles like drug additivity have been explored for small numbers of genes and relatively simple phenotypes like growth rate, drug response principles at the transcriptome scale in diverse cell-populations remain poorly understood.

In order to develop an experimental, conceptual, and mathematical framework for rational programming of immune cell population, we combined large-scale single cell genomics with a physics-inspired mathematical modeling framework to dissect combinatorial drug responses at the transcriptome scale in the human immune system. Using multiplexed single-cell mRNA-seq, we profile the state of 1.5 million human immune cells responding to more than 500 different immunomodulatory drugs alone and in combination. We constructed a mathematical model that predicts the transcriptome scale response of individual cell-types including macrophages, T-cells and B-cells while also predicting how drugs re-balance the relative proportion of cell-types and, thus, alter population structure. The model reveals that combinatorial drug responses can be predicted as emerging as linear and non-linear combinations of single drug responses on individual gene expression programs. Our model quantitatively predicts the progression of a heterogeneous immune cell population between states of activation, inhibition, and rest across drug combinations allowing us to predict optimal drug doses for immune state control. Broadly, our work provides an experimental and mathematical foundation for the design of therapeutic strategies that can rationally reprogram the state of the human immune system.

%A challenge is that core functions of the immune system are executed by populations of specialized cell-types ~\cite{papalexi2018single}. Existing studies of immune cell drug responses have largely applied experimental techniques that measure tens of genes rendering a global analysis of cell state modulation impossible ~\cite{Krutzik2006-hu, Bendall2011-uy}. Rational programming of immune cell state will require transcriptome-scale analysis to determine how drugs and drug combinations can shift immune cells between a spectrum transcriptional states. 

\section*{Results}

\subsection*{T-cell driven immune activation experimental system}

T-cell mediated immune activation is implicated in the pathology and treatment of a variety of human diseases, including cancer and autoimmunity. Cytokines released by activating T-cells can induce a population level switch that drives the activation and signaling of additional cell-types including macrophages leading to the accumulation of inflammatory cell types, and a hyper-active, immune state. Small molecule modulation of T-cell driven immune activation is thus a goal for the treatment of autoimmune disease~\cite{jamilloux2019jak} and for the prevention of cytokine storms that can occur in response to infection also also during chimeric antigen receptor (CAR) T-cell cancer therapy~\cite{fajgenbaum2020cytokine}. 

To probe the impact of small molecules on T-cell driven immune population hyper-activation, we developed an \textit{in vivo} experimental system in which we culture a heterogeneous population of donor derived human immune cells (peripheral blood mononucleated cells, PBMC) and activate the T-cells using tetrameric anti-CD3 and anti-CD28 antibodies~(Fig.~1A). By using a a heterogeneous interacting mixture of T-cells, B-cells, myeloid cells and NK cells, we attempt to mimic the dynamic, multicellular contexts found \textit{in vivo}. In the absence of activating signals, all immune cell types expressed markers of their respective resting state including IL7R (T-cells), KLF2 (T-cells and B-cells), and FCN1, CD14 (monocytes)(SI Fig.~1). 

T-cell activation by the antibody cocktail resulted in global activation of the immune population leading to a network of signaling interactions between T-cells, macrophages, and B-cells. Single-cell time-course transcriptional profiling over 30 hours revealed a cascade of activation and signaling events between all cell-types (Fig.~1B-D) following T-cell activation. Following stimulation, T-cells transitioned from resting to activated states within 2 hours and expressed macrophage-inducing cytokine genes including CCL2, CCL3, CCL4 and IFNG~(Fig.~1C,SI Fig.~1,2). At six hours, monocyte cells transitioned to become activated macrophages and expressed T-cell expansion and recruitment signaling including CXCL9 and IL27 (Fig. 1D), and NK-cells and B-cells transitioned into an activated state at a similar time point. In this way, the immune cell community reached an activated steady state by 20 hours with elevated expression of activation marker genes for each cell type (SI Fig.~1), and we chose this time point to investigate the effects of immunomodulatory drugs. 

%We used the T-cell activation system to analyze drug responses within an interacting immune cell population in which the population responds to an initiating signal, T-cell activation, which drives a cascade of activation events at the cell population level. The system executes a switch in population state, generating activated T-cells, B-cells, macrophages, and NK cells, reminiscent of the T-cell driven inflammatory events that occur in disease processes like cytokine storms and hyper-inflammation. 

\begin{figure}
\vspace*{-1cm}
\includegraphics[width=\textwidth]{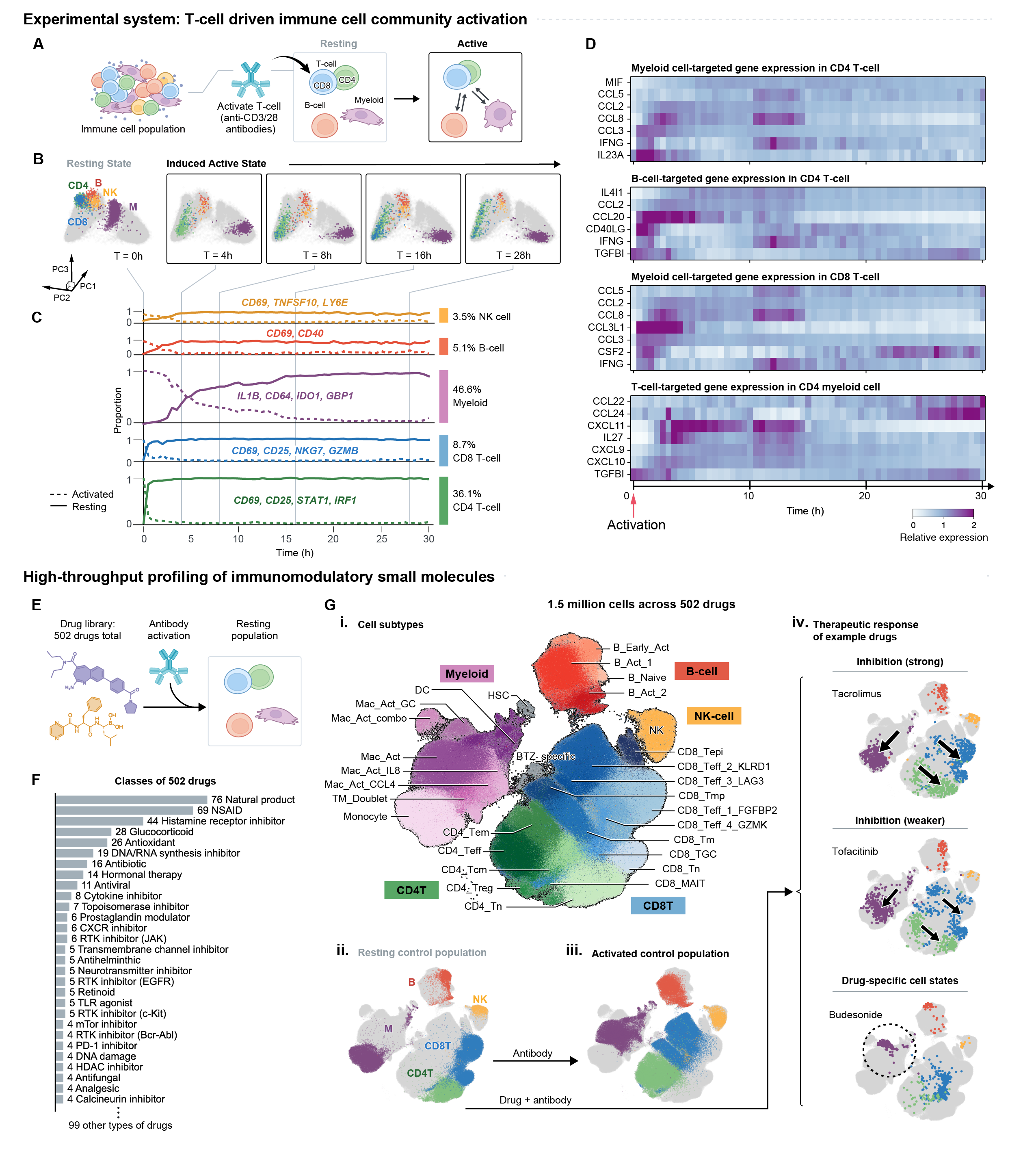}
\label{fig:1}
\vspace*{-0.7cm}
   \caption{\textbf{An \textit{in vitro} T-cell mediated activation system enables high-throughput single-cell profiling of drug responses in an interacting human immune cell population.} (A) Illustration of the T-cell mediated immune activation system. T-cells are specifically activated via anti-CD3/CD28 antibodies, which induce the activation of the entire immune cell population. (B) PCA projections of single-cell transcriptomes at selected timepoints within the 30-hr timecourse experiment. The abundance of each cell-type gradually shifts from resting to activated. (C) Proportion of activated/resting cells in each cell type through time with activation gene markers. T-cells reach activated states first in 2 hrs and myeloid cell activation lasts until 16 hr. (D) Examples of signaling gene expression dynamics in different cell types. Expression is normalized by the mean across the timecourse. (E) Schematic of drug response assay. (F) Histogram of classes of drugs profiled in the experiments. The selected drugs have a variety of biochemical properties and target pathways. (G) UMAP of cell population is used to visualize cell state distributions. (i) Leiden clustering identified 32 cell subtypes and each subtype is curated by marker genes and deferentially expressed genes between subtypes. (ii)(iii) Resting and activated immune population occupy distinct regions on UMAP embedding. (iv) Drugs exhibit different therapeutic effects such as population inhibition and generation of drug-specific cell states not observed in control population.}
\end{figure}

\subsection*{Profiling the cell population response to immunomodulatory small molecules}

We applied the experimental system to measure the impact effect of 502 immunomodulatory small molecules on T-cell driven immune activation~(Fig.~1E). Our drug library contained a diverse set of small molecules targeting pathways including mTor, MAPK, glucocorticoids, JAK/STAT, and histone deacetylases (HDAC)~(Fig.~1F, SI Table). The library contains drugs used for treating auto-immune disease (e.g. Tacrolimus, Budesonide, Tofacitnib) as well as FDA approved anti-cancer drugs (e.g. Bosutinib, Crizotinib). We added each compound at 1{\textmu}M at the same time as anti-CD3/CD28 antibody, and profiled the cell population using single-cell mRNA-seq at 24 hours. We applied lipid-modified oligos~\cite{McGinnis2019-ho} to multiplex and pool samples to increase experimental throughput. In total, we obtained the profile of 1.5 million single cell in resting and activated conditions with over 1,200 total conditions and 31 different immune cell-states of including 4 CD4 T-cell states, 10 CD8 T-cell states, NK cell, 4 B-cell states, and 8 myeloid cell states~(Fig.~1G, SI Fig.~4,5, SI Table). 

%We applied 10x Genomics Cell Ranger 3.1.0 to align the data~\cite{zheng2017massively}, a custom algorithm to demultiplex the data, and scVI to perform leiden clustering and batch correction only for UMAP visualization~\cite{lopez2018deep}. 

Exploratory data analysis revealed that small molecules shifted the structure of the cell population, providing a signature of drug impact. For example, Tacrolimus, a small molecule inhibitor of calcineurin and mTOR protein complexes, blocked the cell state transition from monocyte to activated macrophage (Mac\_Act), as well as naive CD4 T-cell (CD4\_Tn) to Effector CD4 T-cell (CD4\_Teff)~(Fig.~1G iv). Tofacitinib, a JAK/STAT inhibitor, also inhibited both T-cells and macrophage activation but to a lesser extent than Tacrolimus. Budesonide, a glucocorticoid (GC) acting on GC receptor to inhibit inflammatory, also inhibited T-cell and macrophage activation but produced GC-specific macrophage state (Mac\_Act\_GC) and CD8 T-cell state (CD8\_TGC) that were distinct from the resting populations~(Fig.~1G iv). 

\subsection*{Unified network model reproduces cell-state distributions across drug conditions}

\begin{figure}
  \includegraphics[width=\textwidth]{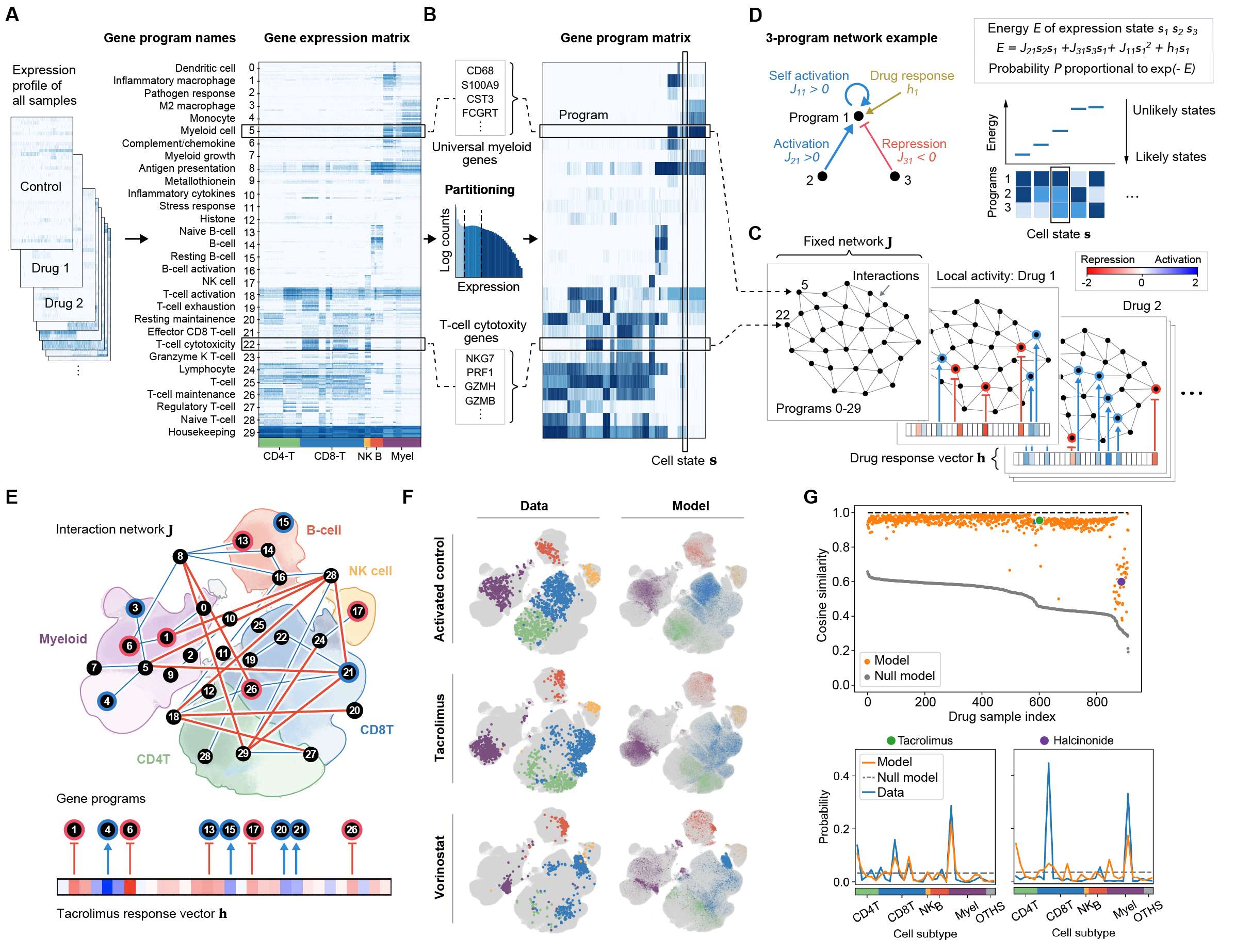}
  \label{fig:2}
   \caption{\textbf{The D-SPIN model quantifies cell population drug response in gene expression programs and reconstructs cell-state distributions.} (A) The filtered gene expression matrix is decomposed into a set of 30 gene programs by linear orthogonal non-negative matrix factorization (B) The decomposition coefficients are discretized into three separate groups (silent, low expression, and high expression) using K-means. (C) Interactions between gene expression programs are inferred from data and generate an interaction matrix $\mathbf{J}$ and drug specific bias vector $\mathbf{h}$. The D-SPIN model predicts the distribution of cell states in each drug condition through a single network $\mathbf{J}$ that is modulated by condition specific $\mathbf{h}$. (D) A schematic illustration of the D-SPIN model shows how interactions between gene programs, encoded in $J_{ij}$, and an external drug specific biasing vector $h$ determine the probability of different cell-states. (E) The specific inferred regulatory network $\mathbf{J}$ rendered by positioning nodes (gene programs) on the UMAP. Each node is positioned on a region of the UMAP where the gene program is highly expressed. Rendering demonstrates that the inferred network decomposes into cell-type specific sub-networks that account for distinct immune cell-types. (F) UMAPs of experimental and model generated cell-state distributions for three conditions showing that the model generated cell-state distribution is qualitatively similar to the experimental data distribution. (G) Top, Quantitative comparison between model and experimental cell-state distribution for all drugs using cosine similarity. The cosine similarity between cell subtype distribution of the model-generated cell state and experimental data is higher than 0.9 for the majority of conditions. For comparison, the null model is a uniform distribution over cell subtypes. Bottom, Examples of model generated cell subtype distribution compared with data. The model and data agree well in Tacrolimus-treated sample. The model fits less well in Halcinonide-treated samples but still captures qualitative features of the distribution.}
\end{figure}

Low-dimensional data embedding approaches like UMAP allow qualitative analysis of cell states and population structures, but do not allow quantitative dissection of drug-response mechanisms, biological interpretation of cell-states, or a strategy for predicting combinatorial drug responses. Therefore, we developed a mathematical modeling framework that allows us to model the interaction of drugs with the immune-cell population as emerging through interactions between drugs and a single, inferred regulatory network. The mathematical modeling framework, D-SPIN (Dimension-reduced Single-cell Perturbation Integration Network), computes the distribution of cell states under different conditions using a minimal probabilistic graphical model~(Fig.~2A-D). The network representation provided by D-SPIN allows us to define drug additivity at the transcriptome scale and also predict combinatorial drug responses by exposing the inferred network to multiple drugs. 

The D-SPIN framework is based on two key ingredients, a low-dimensional representation of transcriptional state and a regulatory-network inference strategy. First, the D-SPIN framework represents the transcriptional state of a cell as a combination of gene programs that contain groups of co-expressed genes 
%that are 
annotated by biological function. Building upon previously published work, we specifically identified a set of 30 gene programs through unsupervised non-negative orthogonal matrix factorization (oNMF) and gene-program annotation~(Fig.~2A, SI Fig.~6). Quantitatively, each gene program in the DSPIN model is set to one of three discrete expression levels corresponding to silent, low and high expression, $(-1,0,1)$~(Fig.~2B). The function of each program was 
manually annotated 
using the gene database~\cite{sherman2022david},
and the single-cell atlas of human immune cells (SI Figs.~7,8)~\cite{dominguez2022cross}. Our decomposition yielded known T-cell and macrophage gene-expression programs including T-cell resting and activation programs, macrophage anti-inflammatory (M2 macrophage) and pathogen-response programs, an antigen-presentation program, and global cell-state programs for B-cells, T-cells, NK-cells, and myeloid cells.

As the second ingredient, D-SPIN constructs a network of regulatory interactions between gene-expression programs and applied drugs through statistical inference. In a given cell-population, combinations of gene programs generate specific immune cell-states including resting and activated T-cells, macrophages, and NK-cells.
D-SPIN computes high- and low-probability gene-program combinations by inferring a regulatory network $\mathbf{J}$ and drug-specific biasing interactions between drugs and the regulatory network encoded in drug-response vectors $h_i$ (Fig.~2D). The regulatory network accounts for positive ($J_{i,j}>0$) and negative $(J_{i j} <0)$ interactions between gene programs. Drugs exert a bias on each gene program influencing it to be on or off through a drug-response vector $\mathbf{h_i}$. Mathematically, an energy function $E(\mathbf{s})$ computes the effective energy of a given gene-program combination by balancing the influence of the regulatory network (encoded in $\mathbf{J}$) and drug-specific inputs (encoded in $h_i$). The resulting probability of a given cell-state $\textbf{s}$ is computed as $P(\mathbf{s}) = \frac{1}{Z} \exp[-E(\mathbf{s};\mathbf{J},\mathbf{h})]$, where 
%$E(s)$, 
\begin{align*}
 E(\mathbf{s};\mathbf{J},\mathbf{h}) = - \sum_{i j}\  J_{i j} s_i   s_j - \sum_i h_i s_i 
\end{align*}
%where E 
is an energy function that weights the impact of the gene-program interaction network, $\mathbf{J}$, and the drug-specific biasing vector, $\mathbf{h_i}$, and $Z$ is a normalizing constant to ensure all probabilities sum to $1$~(SI Text)~\cite{yeo2021generative, weinreb2018fundamental}. The D-SPIN model is motivated by the physics of disordered systems where ``spin-network models" have been applied to capture emergent properties of systems of interacting agents including magnetic materials, networks of neurons, and flocks of interacting birds. 

Conceptually, the D-SPIN model represents cell-states as emerging due to interactions between gene-expression programs and computes the probability of each cell state by weighing gene-program interactions, as quantified in $\mathbf{J}$, with the drug-specific bias. In a 3-program pedagogical example~(Fig.~1C), program 1 receives activation $J_{21} > 0$ from program 2, inhibition $J_{31} < 0$ from program 3, self-activation $J_{11} > 0$ and drug-modulated local activity $h_1$. Repression between programs 1 and 3 
($J_{1,3}<0$)
%leads to 
favors
cell-states with programs 1 and 2 ``on" and with program 3 ``off"
%to be frequent 
(they have
low energy) 
%states 
while 
suppressing
cell-states with program 1 ``on" and program 3 ``on" 
%is a 
(they have
high energy).

%We developed a computational framework for inferring the D-SPIN model from the single-cell mRNA-seq data (Methods,\cite{jiang2019active} ). First, we identified a set of 30 gene programs and annotate each program with a biological functions through unsupervised but curated non-negative orthogonal matrix factorization (oNMF)~(Fig. 2A, SI Fig. 6). Our decomposition yielded known T-cell and macrophage gene expression programs including T-cell rest and activation programs, macrophage anti-inflammatory (M2 macrophage) and pathogen response programs, an antigen presentation program, and global cell-state programs for B-cells, T-cells, NK-Cells, and myeloid cells. Quantiatively, each gene program in the DSPIN model is set to one of three discrete expression levels $(-1,0,1)$ by K-means clustering, representing silent, low expression, and high expression of the gene programs~(Fig. 2B). The function of each program was manually synthesized using the gene database~\cite{sherman2022david} and its expression profile in our data as well as single-cell atlas of human immune cells (SI Fig.7,8)~\cite{dominguez2022cross}.

%Given the thirty gene expression programs, we infer statistical interactions $\mathbf{J}$ between gene programs and drug response vectors that capture known biology of the immune cell population~(Fig. 2D). We found that a single gene program interaction matrix was sufficient to 

When we applied the D-SPIN framework to our data decomposed into 30 gene-expression programs, we found that a single gene-program regulatory network $\textbf{J}$ and a set of 502 drug-specific biasing vectors $\mathbf{h}_i$ were sufficient to describe all 502 drug experiments with low quantitative and qualitative error. The inferred regulatory network can be visualized as an interaction graph (Fig.~2E); the inferred graph contains sub-networks of positive interactions that coincide with major immune cell-types including the T-cell, macrophage, and B-cell states observed in our experimental data (Fig.~2E). In the inferred network, positive couplings occur between closely related gene programs that often appear in the same cell-type; for example, ``Monocyte",``Myeloid growth" and ``M2 macrophage" are all connected to ``Myeloid cell". Negative interactions primarily occur between gene programs expressed in different cell-types, for example, "Antigen presentation" and "T-cell maintenance", as antigen presentation happens in B-cells and myeloid cells
but not T-cells. 

The D-SPIN model reproduces the cell-population structure in both unperturbed and drug-treated conditions. To generate model-predicted cell-populations, we sample from the D-SPIN model using Markov Chain Monte Carlo. The model-generated cell population distribution captured key population-structure transitions including the transition from activated to resting and the emergence of
the
epigenetically disrupted cell-state (CD8\_Tepi) induced by Vorinostat, an HDAC inhibitor used in chemotherapy~(Fig.~2E). Quantitatively, the model and data distributions in $92.4 \%$ of samples the model is {\textgreater}90\% similar to data~(Fig.~2F). There is also a small proportion (4.7\%) of samples 
where the model reconstruction is 
less similar to data with $\sim$70\% similarity. %DAS: not sure what this means
These drugs, for example, glucocorticoids like Halcinonide, substantially alter the cell population structure and create cell-states not found in control populations. For these drugs, quantitative accuracy is worse,
but the D-SPIN model generates distributions with the correct trends and therefore can identify these drugs
as
being effective and reproduce their population-response signature. The D-SPIN model is simple and interpretable by using only 30 parameters for each 
to summarize the 
drug-condition
response that contains 
of
30,000 distinct genes per cell
across
more than 1,000 single cells.

\subsection*{Classification of compounds into phenotypic classes using the drug-response vector} 

The drug-response vectors, $\mathbf{h_i}$, identified by the D-SPIN model are compact, interpretable, and generative signatures of drug impact on the gene-regulatory networks controlling the activation of the immune-cell population. We used the drug-response vectors to classify all drugs into a set of seven classes using $K$-means clustering: strong inhibitor, weak inhibitor I, weak inhibitor II, glucocorticoid, M1 macrophage inducer, epigenetic regulator, and toxicant (Fig.~3A, SI Fig.~10). Each inhibitor class has a distinct immunomodulatory cell-population response signature.

\begin{figure}
  \includegraphics[width=\textwidth]{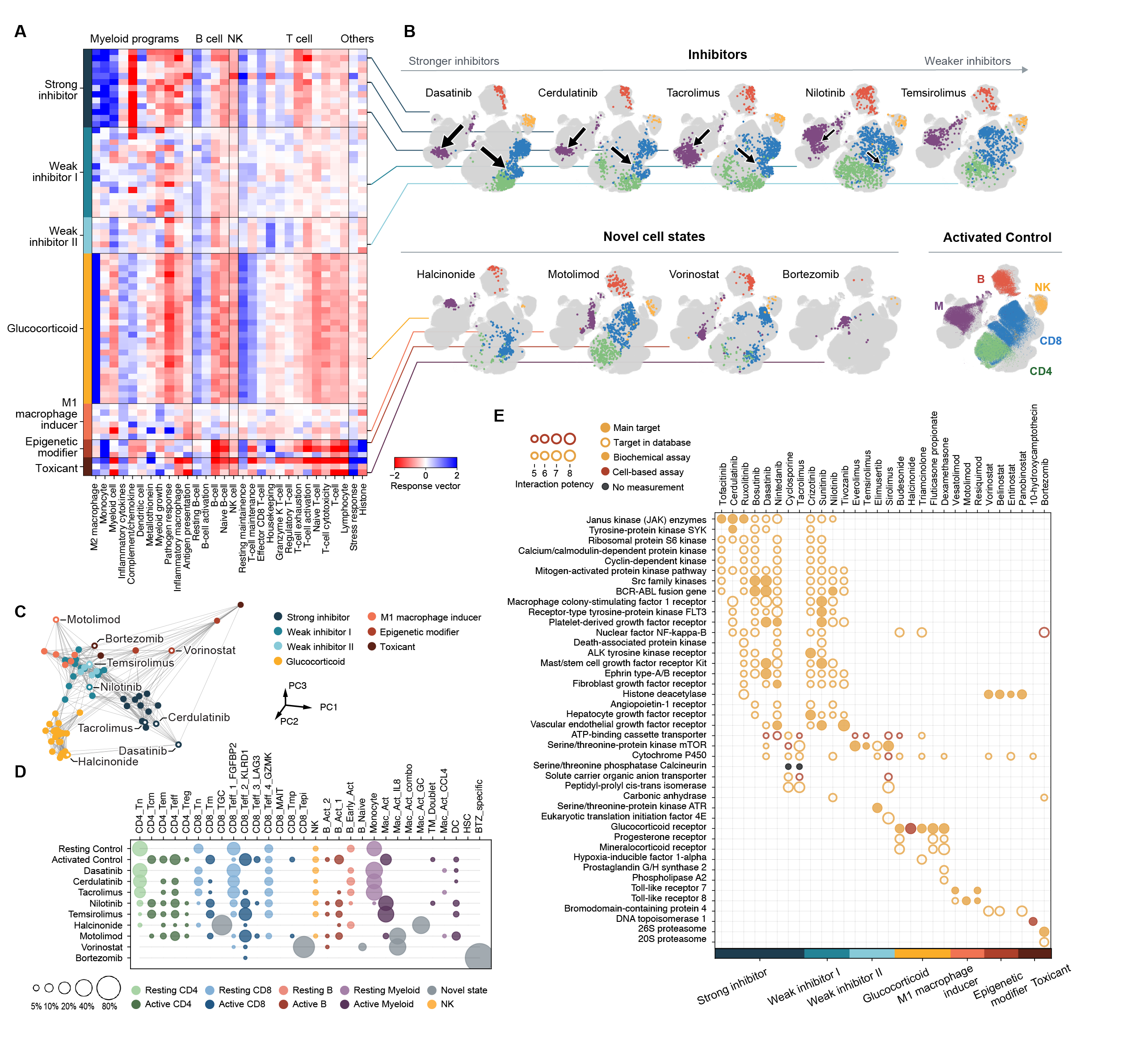}
  \label{fig:3}
   \caption{\textbf{Drug-response classes identified by D-SPIN reveal inhibitors with various strengths and different drug-induced novel cell-states.} (A) Response vectors and drug classes for identified effective drugs. Gene programs are grouped by major cell types where they are present. (B) UMAPs of selected drugs from different classes. We find immune inhibitors with a spectrum of strengths, and drugs producing novel cell-states. (C) 
   %PCA 
   Leading principal components
   of response vectors of identified effective drugs visualize the relationship between different drug groups. (D) Bubble plot of cell-subtype distributions of control and selected drug examples. 
   With decreasing inhibitor strength, %DAS: how do I read inhibitor strength in the panel?
   the proportion of the activated immune-cell population (deep colors) gradually increase. Some drugs produce cell-states different from both resting and activated control populations. (E) Targets and mechanisms %DAS: mention which is which in the panel (rows vs. columns)
   for selected drugs. The drugs acting on a variety of molecular targets and pathways.} %DAS: mention potency and what it means?
\end{figure}

Three of the seven classes %DAS: "category" and "class" mean the same thing? If so, be consistent and stick with one. 
include drugs that inhibit activation of the immune-cell population (Fig.~3B), including drugs we classified as strong inhibitor, weak inhibitor I, and weak inhibitor II. Across these drug classes, analysis of cell-population fractions and UMAP visualization indicated that these drugs block the transition of T-, B- and myeloid cell types to the activated state, shifting the balance of the cell population to resting cell-states and acting with a spectrum of inhibitor strengths (Fig.~3B-D). Very strong inhibitors, including the immunosupressive drug Tacrolimus and the cancer drug Dasatinib, completely block the activation of T-cells and myeloid cells, inducing a cell population with a similar cell-state profile to unstimulated PBMC cells (Fig.~3D). Weak inhibitors, such as Temsirolimus, only slightly bias the cell population towards the resting direction compared with the control population.

The individual drugs that fall into the inhibitor drug classes act on a variety of molecular targets and signaling pathways ~\cite{skuta2017probes}~(Fig.~3E). The strong-inhibitor class targets include JAK (Tofacitinib, Cerdulatinib, Ruxolitinib), BCR-ABL and Src family kinase (Bosutinib, Dasatinib), and calcineurin (Cyclosporine, Tacrolimus), which are all pathways downstream of T-cell receptor signaling. The drugs in the weak-inhibitor I class target another group of receptors including ALK (Crizotinib), FLT3 (Sunitinib) and various growth-factor receptors (Crizotinib, Sunitinib, Nilotinib, Tivozanib). The drugs in the weak-inhibitor II class target protein kinases including mTOR (Everolimus, Temsirolimus, Sirolimus) and ATR (Elimusertib). Broadly, the analysis suggests that inhibition can be achieved via a range of distinct biochemical pathways and mechanisms. 

More detailed comparison between the effects of drugs with the same primary target demonstrates that the single-cell assay and D-SPIN model can identify phenotypic patterns that capture potential underlying differences in the biochemical mechanism of drug action~(Fig.~3E). A specific set of BCR-ABL kinase inhibitors (Bosutinib, Dasatinib) developed for the treatment of leukemia generated strong inhibition, while another set of BCR-ABL inhibitors, including Nilotinib, fall into the weak-inhibitor class. Biochemically, the BCR-ABL strong inhibitors can be distinguished from the BCR-ABL weak inhibitors by their increased Lck/Src kinase specificity~(Fig.~3E). Lck kinase is a central signaling node directly downstream of T-cell receptor activation. Similarly, the JAK-kinase inhibitor Cerdulatinib generated a stronger inhibitory response than the FDA approved anti-inflammatory JAK-kinase inhibitor Tofacitinib. While both drugs have JAK-family kinases as their primary target, Cerdulatinib also targets the kinase SYK. Therefore, the response vectors generated by D-SPIN allowed us to identify potential signaling nodes that differentiate responses 
%between 
from
related drugs. 

Beyond the strong and weak inhibitors, the other identified classes include drugs that induced novel cell-states that are different from 
%the transcriptional states 
those
found in both the resting and activated control samples. Glucocorticoids, a class of widely used anti-inflammatory drugs, not only inhibit T-cell activation but also produce an M2 macrophage state that is related to tissue repair~\cite{desgeorges2019glucocorticoids}. The M1 macrophage inducer class contains TLR7 agonists (Vesatolimod, Resiquimod) and TLR8 agonists (Motolimod, Resiquimod), and produced macrophage states related to pathogen responses. The epigenetic-modifier class consists of HDAC inhibitors which generated an epigenetically disrupted T-cell state (CD8\_Tepi) that expressed high levels of histone-related genes. The toxicant class caused cell death and high expression of stress-response-related genes. This class contains HDAC inhibitor (Panobinostat), proteasome inhibitor (Bortezomib) and DNA topoisomerase inhibitor (10-Hydroxycamptothecin). 

%Together, our analysis identified classes of immunomodulatory drugs by characterizing population structure and drug-response vectors, with distinct drug-action mechanisms resulting in similar population structures. Biologically, we find that drug effects on T-cell and B-cells 
%are mostly tune the balance between activated and resting, while drugs impact myeloid cells by switching cells between a group of different myeloid states. 

\subsection*{Drug combinations generate novel cell-states explained by an additive response model}

Practically, immunomodulatory drugs are often used in combination to modulate the immune system, and the rational design of drug combinations that achieve clinically important immune states of repression and activation remains a major challenge. More conceptually, drug combinations have been used extensively to dissect cellular regulatory networks. Drugs that combine in additive manner typically act on independent regulatory pathways. Non-additive drug interactions can identify pathways that interact with one another to control cell-state. Therefore, we applied the D-SPIN model to predict and analyze the impact of drug combinations on the experimental immune-cell population. 

Theoretically, the D-SPIN model provides a natural mathematical framework for identification of additive and non-additive drug responses at the transcriptome scale. Specifically, the D-SPIN energy function can be derived from biochemical regulatory network models where the $\mathbf{J}$ matrix corresponds to regulatory network interactions between gene regulators, and the drug-response vectors $\mathbf{h_i}$ correspond to network inputs provided by drugs. In the D-SPIN model, drug inputs that are independent of one another will simply lead to an additive drug-response vector $h_{1+2} = h_1 + h_2$ while drug inputs that interact with one another lead to deviations from additivity. In this way, 
given the assumption of additivity
the D-SPIN model allows us to predict the cell population structure induced by drug combinations. 
Thus, the model provides a mathematical framework for 
examining drug interactions globally 
by asking if responses match or deviate from the additive prediction. 

To probe experimentally the impact of drug combinations at the transcriptome scale, we selected 12 drugs from the different phenotypic drug classes identified in Figure 3 and performed all possible pair-wise combination experiments. At the population level, drug combinations both rebalanced existing cell-states and also generated novel drug-combination-specific cell-states. For example, when we combined strong inhibitors like Dasatinib and Tacrolimus, the resulting cell populations exhibited inhibition, with populations dominated by ``resting-like” cell-states (Fig.~4B, SI). Similarly, when drugs from the toxicant or HDAC-inhibitor drug classes were combined with drugs from all other classes (Fig.~4B) the resulting cell-populations resembled toxicant or HDAC-inhibitor cell populations generated by single drugs. When drugs that induced novel cell-states were combined with drugs from other classes, the drugs generated cell-types not observed in single-drug 
%response 
experiments. Halcinonide combined with strong inhibitors (Bosutinib, Dasatinib, Cerdulatinib, and Tacrolimus) produced a novel macrophage state that was not observed when adding a single drug with the activating antibody (Fig.~4B). The combination between Motolimod and strong inhibitors yielded a hybrid population structure 
%where we observe 
with
a Motolimod-induced macrophage state and resting-like T-cells (Fig.~4B).

\begin{figure}
  \includegraphics[width=\textwidth]{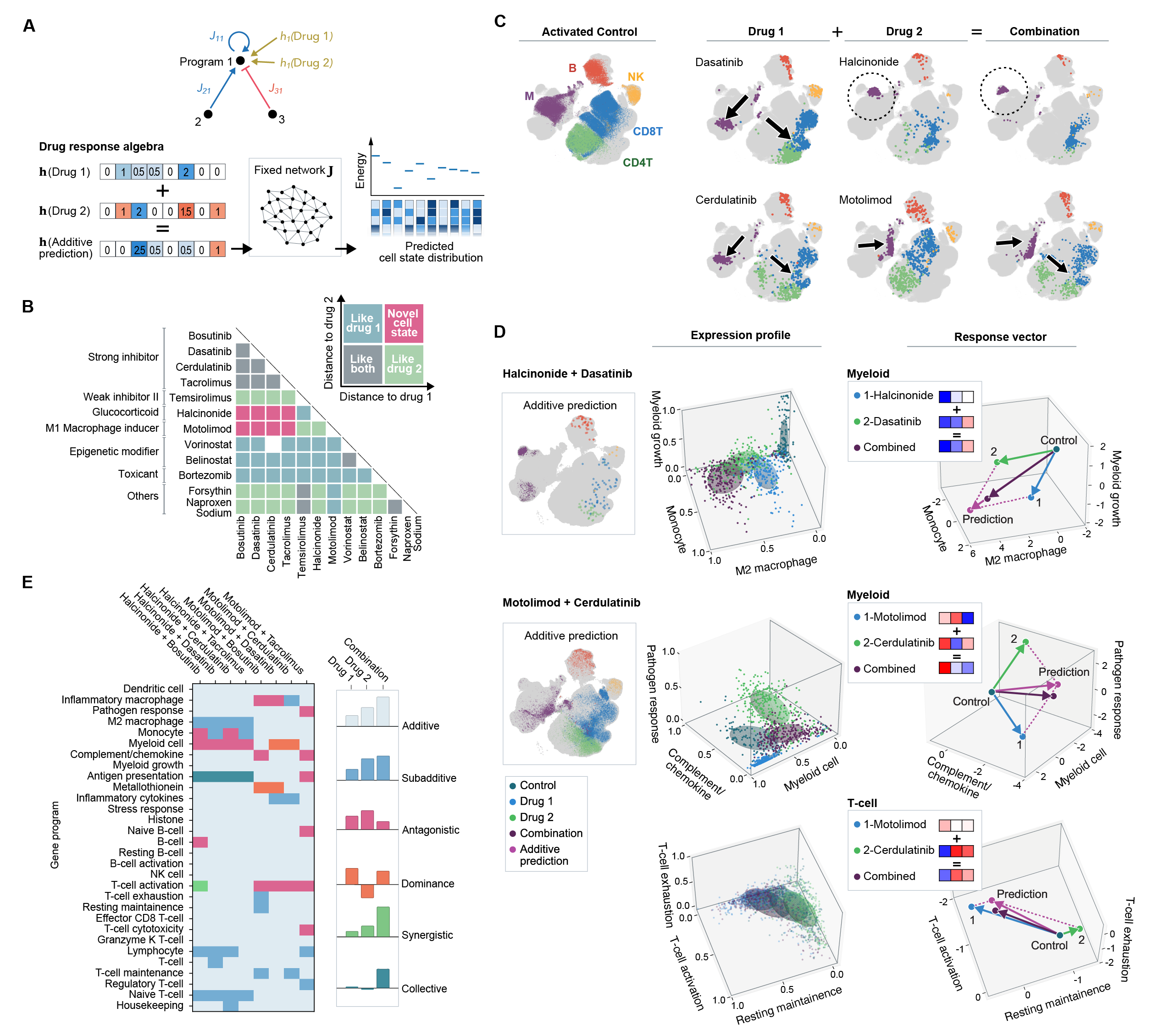}
  \label{fig:4}
   \caption{\textbf{Drug combinations produce novel cell-states that can be understood by additivity in the model.} (A) Schematic of additive drug effects and additive prediction of cell-state distribution by D-SPIN. (B) Classes of selected drugs and drug-combination outcomes. The majority of drug combinations have similar effects as one of the single drugs, except combinations 
   involving Halcinonide or Motolimod with strong inhibitors. (C) Examples of novel cell-state distributions produced by combinations. Halcinonide plus Dasatinib generate 
   %a 
   novel myeloid states. Motolimod plus Cerdulatinib generate inhibited T-cell states with Motolimod-like myeloid states. (D)(i) %DAS: no (i)/(ii) labels in plot; replace with top/bottom?
   Additive model predicts the emergence of novel myeloid state in Halcinonide plus Dasatinib treatments. 
   Scatters of T- and B-cells are enlarged %DAS: not quite sure what this means; rephrase?
   for visualization as the predicted distribution is mostly dominated by myeloid cells.
   Expression profiles of drug combination exhibit the impact of both single drugs. The fitted response vector agrees well with the additive prediction on these 3 gene programs. (ii) Additive model predicts the hybrid population structure of Motolimod plus Cerdulatinib. The expression profile and response vector on key myeloid and T-cell programs display a similar additive relationship. (E) 
   %DAS: the actual data is something else: the additivity/subadditivity/etc, shown for drug combinations that somewhere show non-additivity.  And it appears to show all programs, not just those with non-additivity (some rows are all additive).  
   Non-additive categories and types of interactions for gene programs in drug combinations generating novel cell states.}
\end{figure}

We compared the experimentally measured cell-population to additive predictions of the D-SPIN model. Specifically, 
%we use 
using
the program interaction network $\mathbf{J}$ 
%that we 
inferred 
%based on 
from
single-drug experiments,
%. Then, 
we compute the predicted distribution of cell-states for an additive drug combination as 
\begin{align*}
    &P(\mathbf{s}) = \frac{1}{Z} \exp[-E(\mathbf{s};\mathbf{J},\mathbf{h}(\text{\small Additive}))], \quad \mathbf{h}(\text{\small Additive}) = \mathbf{h}(\text{\small Drug 1}) + \mathbf{h}(\text{\small Drug 2})
\end{align*} %DAS: can you drop this equation and just refer to your original displayed equation for E, then specify inline the additive form of the h?
where $ \mathbf{h}(\text{\small Drug 1})$ and $\mathbf{h}(\text{\small Drug 2})$ are the drug-response vectors inferred from the single-drug experiments. For drug-combination experiments, the additive drug-response vector, generated through simple addition of single-drug-response vectors, can be directly compared to a combinatorial drug-response vector inferred directly from the drug-combination 
%data. 
experiments.

Across drug combinations, the additive model 
%generated 
predicted 
cell 
populations that captured key features of the 
%empirical data. 
experiments.
Specifically, combinations of inhibitory drugs generated inhibited populations in model and 
%data 
experiment 
(SI). 
%In the case of 
For
toxicants, the individual toxicant vector 
%was able to dominant 
dominated
the combination, 
%generating a vector that produced
producing
a toxicant-like cell population (SI). In these cases the additive prediction and 
%empirical 
experimental
population distribution show strong agreement both in UMAP and cosine-similarity calculations (SI). 

The novel-cell-state drugs provide an interesting test of the additive model because drug combinations generate cell-states that occupy ``new'' regions of the UMAP embedding that are not occupied for single drugs (Fig.~4C, dotted circle, top) as well as 
novel population structures %DAS: rephrase? Not sure what this means.
(Fig.~4C, bottom). 
%In the case of 
For
Halcinonide 
%and 
plus
Dasatinib, the additive model 
%was able to 
successfully
predicted the emergence of the novel combination-specific cell-state and provided a mathematical framework to interpret 
%these states 
this state
in terms of the transcriptional states of single-drug responses~(Fig.~4D). The novel myeloid state generated by Halcinonide plus Dasatinib is distinct from the response 
%of 
to
either drug alone, but can be understood at the gene-program level using the additive response: Halcinonide-induced macrophages 
%were characterized by the 
had
high expression of the ``M2 macrophage" gene program including anti-inflammatory genes CD163, MS4A6A and VSIG4, while Dasatinib promoted the expression of the "Monocyte" program including endosome-biogenesis-related genes APOE and CD9. Together, Dasatinib 
%and 
plus
Halcinonide generates a macrophage population that is a super-position of the single-drug gene-expression states~(Fig.~4D): The drug combination activated both ``M2 macrophage" and ``Monocyte" programs. For 3 selected key gene programs in myeloid cells, the shift of expression distribution from Halcinonide alone to the drug combination is similar to the shift from control to Dasatinib alone, indicating the regulatory effect on these gene programs of the drug combination is the addition of two single drugs. 
%DAS: isn't the previous sentence just a re-statement of the next sentence (and the next sentence is the common way you talk elswehere about additivity) ?  
Moreover, the response vector of the drug combination on the 3 programs agrees quantitatively with the algebraic sum of the vectors of the two single drugs (Fig.~4D). 

As a second example, the TLR8 agonist Motolimod alone generates a cell population with activated macrophages that have high expression of the ``Pathogen response" program 
%with 
including
host-defense signaling genes like CXCL1, IL6 and IL8, and also promotes T-cell activation through weak inhibition of the ``Resting maintenance" naive T-cell program. Alternately, the strong JAK inhibitor Cerdulatinib induced a cell population with resting-like T-cells and resting macrophages. Together, the drug combination induced a mixed or frustrated cell-population that contains macrophages with an intermediate transcriptional state with decreased expression of the "Pathogen response" program and resting-like T-cells. The additive prediction in D-SPIN is consistent with the cell-state distributions in the experiments. The relative magnitudes of the response vector lead to a hybrid population of Motolimod-like macrophages and resting-like T-cells~(Fig.~4D). The expression profile on 3 key programs for myeloid and T-cell each revealed that the drug combination is the superposition of the effects of two single drugs alone. 
%DAS: again, the previous sentence sounds like a restatement of the next sentence (which is the more standard way you have been describing additivity throughout).
We again found quantitative agreement for the selected programs between the additive prediction and the response vector of the drug combination. 

Across drug combinations, the additive model was sufficient to predict key features of the drug-combination-induced %DAS: could probably drop "drug" in these phrases
cell populations including the emergence of novel, drug-combination-specific cell states. However, the additive model diverged from the data due to non-additive drug interactions that could be identified in specific gene-expression programs. We computed how each program contributes to the deviation between the additive prediction and the experiments, and classify each gene program as additive, synergistic, antagonistic, etc. following the drug-epistasis literature~(Fig.~4E). As an example, 
%in the case of 
for
Dasatinib and Halcinonide, we identified significant antagonistic interaction between the drugs on the ``Myeloid cell" program, where both drugs alone activate this program but their combination produces inhibition. Biologically, we observed the most significant non-additive effects for myeloid gene-expression programs where drug combinations allow the cell population to access novel cell-states not observed for 
single drug. %DAS: I think you've called this "drugs alone" in the past?

\subsection*{Drug-combination dosing provides access to spectrum of novel cell states}

\begin{figure}
\includegraphics[width=\textwidth]{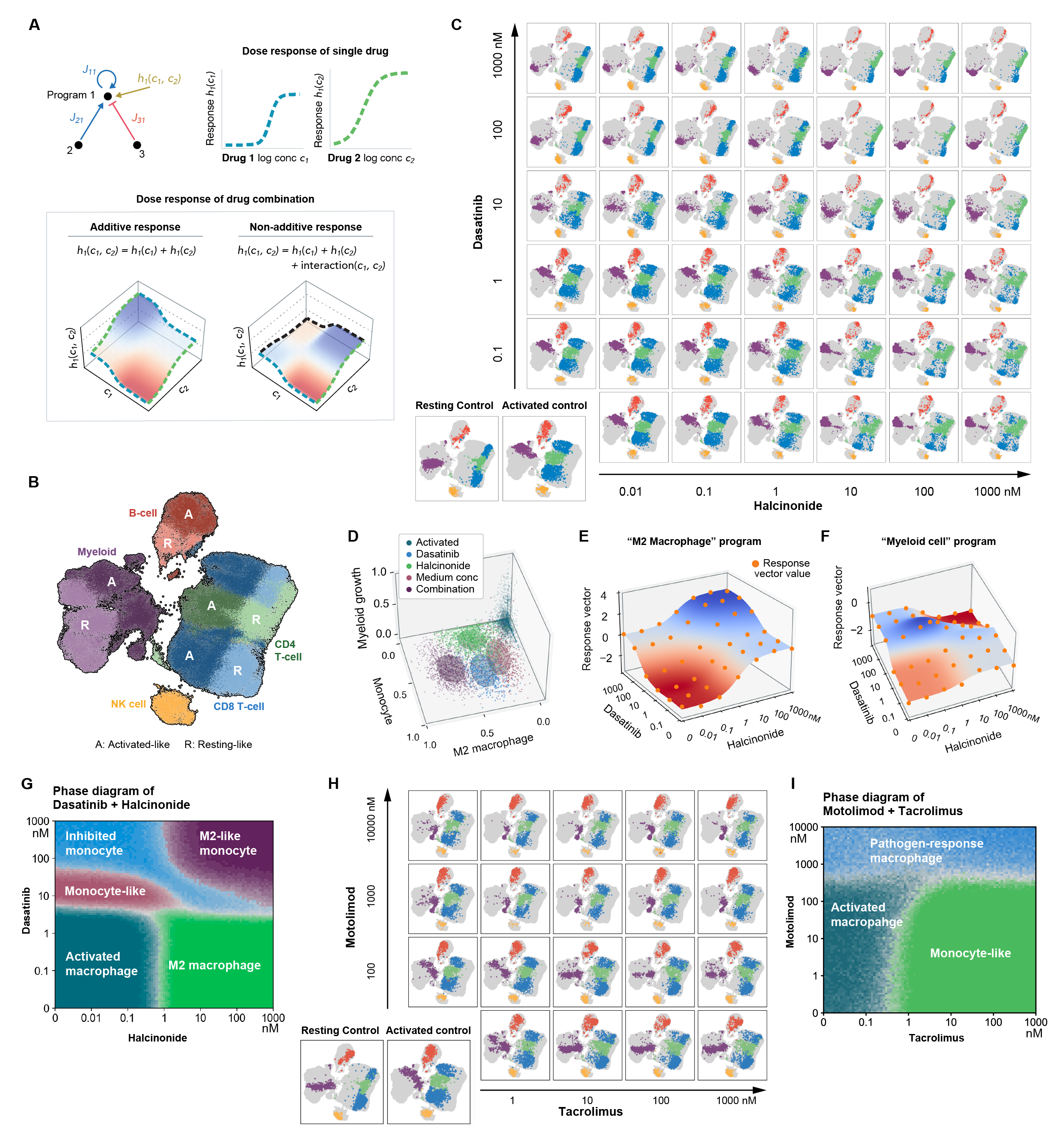}
\label{fig:5}
   \caption{\textbf{Drug-combination dosing provides access to continuous spectrum of novel cell states that can be depicted by gradual changes in the model response vectors.} (A)Schematic of additive drug response and non-additive drug response in D-SPIN for dosage titration experiments.  
   (B) UMAPs and cell-type labels for all combination dosing conditions. (C) UMAPs of the cell population treated by different dosings of Halcinonide plus Dasatinib. Cell population gradually shifts with the increase of drug concentration. (D) UMAP and expression profile on key myeloid programs for control, Halcinonide, Dasatinib, combination and the resting-like myeloid population produced by medium doses of each drug. 
   %DAS: I think D in figure is missing any description in caption
   (E) Example gene program that follows additivity. (F) Example gene program that deviates from additivity. Both single drugs produce activation, but their combination produces inhibition. %DAS: are D and E in the caption E and F in the figure?
   (G) Model-computed phase diagram of myeloid population as function of Dasatinib and Halcinonide dosing. Myeloid cell-state transits from resting-like to Dasatinib-induced to combination-induced states as Halcinonide concentration increases. %DAS: G in figure?
   (H) Subset of UMAPs of the cell population treated by Motolimod plus Tacrolimus. (I) Model-computed phase diagram of myeloid population as function of Motolimod and Tacrolimus dosing.}
\end{figure}

Dose titrations can be applied to analyze non-additive drug interactions by asking how the response of a cell population to a single drug changes as a second drug is added at increasing concentrations. Titrations can, therefore, reveal deviations from additivity by showing how the presence of a second drug alters the form of a drug response (Fig.~5A). Further, we wanted to understand if drug titrations could be applied to program the state of macrophage cells and the immune population structure more broadly. 

Deviations from additivity can be analyzed within the D-SPIN framework by asking how drug-response vectors compare to additive predictions across combinatorial drug doses (Fig.~5A). Experimentally, we characterized the non-additive effects of drug combinations by selecting two drug pairs, Halcinonide plus Dasatinib and Motolimod plus Tacrolimus. We performed combinatorial dose titrations spanning the concentration space from 0.01nM to 10,000nM, selected based on reported IC50s for all drugs (Fig.~5B-H). We performed experiments at logarithmically spaced drug concentrations profiling $42$ concentration pairs for Halcinonide and Dastanib and $36$ for Motolimod and Tacrolimus (Fig.~5B,H). Comparison of experiment with the D-SPIN model allowed us to localize non-additive effects to specific gene-expression programs. 

Broadly, we found that the dose titration of Halcinonide plus Dasatinib produced a continuous spectrum of macrophage and T-cell states~(Fig.~5C) and population structures. For both myeloid cells and T-cells, the cell population gradually shifts to more inhibited states with increasing concentration of either drug. At low concentrations of both drugs, the cell population closely resembles the activated cell population (Fig.~5C). At medium concentrations of Dasatinib and Halcinonide, the myeloid population is similar to the resting control, occupying a distinct region on UMAP and key gene-expression programs~(Fig.~5C,D). The resting-like cell-state cannot be generated by either Halcinonide or Dasatinib alone. At high concentrations of both drugs, the cell population shifts to a hyper-inhibited state with the macrophage cells activating the glucocorticoid-specific gene-expression program that contains anti-inflammatory genes TGF-beta and CD163, genes that play an important role in anti-inflammatory M2 macrophages observed in both cancer and auto-immune disease responses. By modulating drug dose, the cell-population sweeps across a curved manifold in transcriptional space, transitioning continuously from activated to resting to hyper-inhibited (SI).

To analyze the interaction of Halcinonide and Dasatinib on non-additive programs, we compared across dosages the drug-combination response vector with the sum of single drugs. Both Halcinonide and Dasatinib activate the ``M2 Macrophage" program, and the combination is well-described by the sum of two sigmoid functions of single-drug doses, except at high concentrations of both drugs where the program exhibits saturation~(Fig.~5E). Thus the M2 macrophage program only exhibits sub-additivity at high concentrations. As a second example of non-additivity, the ``Myeloid cell" program is a core gene program generally expressed in myeloid cells, therefore representing the abundance of myeloid cells in the cell population. Both Halcinonide and Dasatinib slightly activate the program when acting alone, but their combination at high dosage is repressive~(Fig.~5F), which is non-additive and 
non-monotonic. %DAS non-monotonic in what? If I take a slice through the data at constant [Das] or constant [Halc], it looks pretty monotonic.  
%DAS: check panel refs, as the figure seems to have added D, shifting all later panels by one.
The non-additive effect explains the deviation between cell-type proportions in the additive model and experiment. %DAS: I don't think it "explains" it. "The non-additive effect" is a shorthand label for "the deviation b/w cell-type proportions in additive model and experiment". They're two different ways of saying the same thing. 
The additive model predicts high doses of Halcinonide plus Dasatinib will strongly activate the universal myeloid program, leading to high macrophage abundance~(Fig.~4D). In fact, non-additive interactions between the drugs inhibit the program and decrease abundance of macrophages relative to the additive model.

We refined the D-SPIN model to include both dosing and non-additive drug responses and trained the model on the dosage-combination data (SI) achieving a high quantitative accuracy as measured by cosine similarity (SI). The resulting model contains both additive and non-additive programs. For the additive programs the response vector is the algebraic sum of two sigmoid dosing functions. For the non-additive programs, we append two extra terms to the additive model to accommodate non-additive effects~(SI).
For example, for the non-additive ``Myeloid cell", both single drugs produce activation but their combination produces inhibition~(Fig.~5F). The refined D-SPIN model, accounting for additive and non-additive drug responses, yielded a quantitatively accurate model of the cell population. 

Using the model, we can predict the structure of the cell population as a function of drug concentrations to establish a phase diagram of the cell population across dose titrations of two drugs. The resulting phase diagram is consistent with the drug-titration experiments and also allowed us to interpolate unsampled regions of the concentration space thereby identifying potential phase boundaries between immune-population states~(Fig.~5G). The phase diagram shows that, at low levels (\textless 10nM) of Dasatinib, the myeloid population transits at 1nM of Halcinonide from activated to the M2 macrophage state. At high levels (\textgreater 100nM) of Dasatinib, the myeloid population transits from Dasatinib-induced inhibition to the combination state. More complex phenomena occur at moderate levels of Dasatinib, where with increasing dose of Halcinonide the myeloid population transits from the resting-like state to the Dasatinib-induced state and the combination state, and the transition depends on the concentration of both drugs. Specifically, the likelihood of a resting-like population is maximized at 10nM Dasatinib and 1nM Halcinonide. In this way, the quantitative model allows prediction of optimal drug doses to achieve immune-states of interest. 

We performed a similar dose-titration analysis for Motolimod plus Tacrolimus. In this second case, we refined the additive model by introducing drug interactions for specific programs (SI). The resulting model reveals that the Motolimod-specific myeloid state is dominant over the Tacrolimus-induced states~(Fig.~5H,I). We compute the phase diagram of Motolimod plus Tacrolimus using the same model adjustment. At low concentration (\textless 1,000nM) of Motolimod, the myeloid cell-state transits with the increase of Tacrolimus concentration from activated to a Tacrolimus-induced resting-like state, and the transition concentration increases with Motolimod concentration. At high concentration (\textgreater 1,000nM) of Motolimod, the myeloid cells exhibit the Motolimod-induced state regardless of Tacrolimus concentration. %DAS: suggest replacing "BLAH concentration" with "[BLAH]" throughout. Will save a lot of words.  

In conclusion, the dosing of drug combinations provides access to a spectrum of cell states quantified in the D-SPIN model by continuous change of response vectors. By parameterizing non-additive drug interactions in specific gene-expression programs, we obtain a quantitative model of the cell population structure as a function of combinatorial drug doses. 
The model allows us to compute optimal drug concentrations to achieve immune-states of interest. %DAS: virtually an identical sentence appears 1.5 paras ago
In this way, we capture the linear and non-linear algebra of combinatorial drug responses within a single predictive mathematical framework. Biologically, the dosing experiments suggest tunable modulation of immune-cell population. The phase diagram computed by the D-SPIN model provides guidance for rational design of new cell-types and new therapeutic strategies. 

\section*{Discussion}

In this work, we demonstrate that certain drug combinations can induce novel cell-states through additive regulation mediated by individual drugs, and dosing combinations can yield a spectrum of cell states. Using single-cell profiling and our D-SPIN model, we identify the transcriptional effects of a variety of immunomodulatory drugs on immune-cell populations, and reveal that similar population responses can be achieved by diverse molecular mechanisms. We find that novel cell-states arise in the presence of specific drug combinations due to the addition of the regulatory effects of single drugs on specific gene-expression programs. We further generate a spectrum of cell states and population structures using drug dosage titration and compute the phase diagrams of cell-states as functions of drug dosages.

Our work establishes an interpretable model framework for transcriptome-scale perturbation datasets. The D-SPIN framework is a maximum-entropy modeling framework adapted from analysis of systems in condensed-matter physics. The framework provides a low-dimensional, compact yet predictive framework for analyzing population-level effects in single-cell data through a regulatory-network model. Linear modulation of parameters in our model enables the use of simple algebraic operations for comparison between conditions, predicting combinatorial response, and interpolating between dosages.

The additivity of drug-combination responses at the gene-program level provides a simple principle for interpreting and predicting the effect of drug combinations. If additivity extends beyond our studied system, our model could facilitate the programming of target states in cell populations. Even though we have shown additivity alone is not able to accurately predict cell-population structure, it could still serve as a guide to narrow down potential drug-combination targets for therapeutic objectives. Additivity could arise from the modularity of gene-regulatory circuits, such that different pathways impact gene expression levels independently. Further work is needed to reveal conditions where additivity holds or breaks down. 

Biologically, we find that drugs and drug combinations tend to simply shift the balance between T-cell and B-cell populations while inducing combinatorial transcriptional sub-states in macrophages. In our experiments, T-cells and B-cells exhibit switch-like transitions between resting and active states. Conversely, macrophages can be tuned between 
distinct and combinatorial states %DAS: rephrase? I'm not clear what this means.
of gene expression that are consistent with functionally pro- and anti-inflammatory signaling states. Similarly, the non-additive drug interactions that we detect occur primarily in macrophage gene-expression programs, suggesting that these programs are controlled by multiple drug-modulatable %DAS: modulatable? In vivo the drug is not normally present ...
signaling pathways. In fact, previous research has shown that the glucocorticoid-receptor pathway which is impacted by drugs like Halcinonide also integrates signals from a variety of kinase signaling pathways that might be impacted by Dasatinib~\cite{weikum2017glucocorticoid}. Together, our results suggest that macrophages can generate a broad spectrum of gene-expression states and that gene programs are potentially controlled by densely interacting signaling pathways. 

Mechanistically, our computed cell-state phase diagrams, which change with drug-combination dosage, provide potentially useful biomedical insight. Different dosage combinations between the BCR-ABL kinase-inhibitor Dasatinib and the glucocorticoid Halcinonide produce a wide range of myeloid cell states, with different level of inhibition and glucocorticoid gene-program activation, providing a diverse range of inflammatory conditions for treating patients. Similar approaches could be applied to other drug combinations to expand the scope of accessible cell-states, making progress towards the targeted design of cell-population states.

\section*{Acknowledgements}

The single-cell profiling experiments were performed at the Beckman Institute Single-cell Profiling and Engineering Center (SPEC). Sequencing was performed at the UCSF CAT, supported by UCSF PBBR, RRP IMIA, and NIH 1S10OD028511-01 grants. We acknowledge Dr.\ Guy Riddihough for editorial assistance with the manuscript. DAS acknowledge the support of a Tier-II Canada Research Chair.

 \newpage
 \bibliographystyle{aiaa}
 \bibliography{reference}
\end{document}